\documentclass[aps,prl,reprint,superscriptaddress]{revtex4-1}

\usepackage{graphicx}
\usepackage{hyperref}
\usepackage{color}
\usepackage[normalem]{ulem}
\usepackage{lineno}
\usepackage{amssymb}

\begin{document}

\title{The effect of $f$-$c$ hybridization on the $\gamma\rightarrow\alpha$ phase transition of cerium studied by lanthanum doping}

\author{Yong-Huan Wang$^\star$}
\affiliation{Science and Technology on Surface Physics and Chemistry Laboratory, Jiangyou 621908, Sichuan, China}

\author{Yun Zhang$^\star$}
\affiliation{Science and Technology on Surface Physics and Chemistry Laboratory, Jiangyou 621908, Sichuan, China}

\author{Yu Liu$^\star$}
\affiliation{Laboratory of Computational Physics, Institute of Applied Physics and Computational Mathematics, Beijing 100088, China}
\affiliation{Software Center for High Performance Numerical Simulation,
China Academy of Engineering Physics, Beijing 100088, China}

\author{Xiao Tan}
\affiliation{Science and Technology on Surface Physics and Chemistry Laboratory, Jiangyou 621908, Sichuan, China}

\author{Ce Ma}
\affiliation{Science and Technology on Surface Physics and Chemistry Laboratory, Jiangyou 621908, Sichuan, China}

\author{Yue-Chao Wang}
\affiliation{Laboratory of Computational Physics, Institute of Applied Physics and Computational Mathematics, Beijing 100088, China}

\author{Qiang Zhang}
\affiliation{Science and Technology on Surface Physics and Chemistry Laboratory, Jiangyou 621908, Sichuan, China}

\author{Deng-Peng Yuan}
\affiliation{Science and Technology on Surface Physics and Chemistry Laboratory, Jiangyou 621908, Sichuan, China}

\author{Dan Jian}
\affiliation{Science and Technology on Surface Physics and Chemistry Laboratory, Jiangyou 621908, Sichuan, China}

\author{Jian Wu}
\affiliation{Science and Technology on Surface Physics and Chemistry Laboratory, Jiangyou 621908, Sichuan, China}

\author{Chao Lai}
\affiliation{Science and Technology on Surface Physics and Chemistry Laboratory, Jiangyou 621908, Sichuan, China}

\author{Xi-Yang Wang}
\affiliation{Science and Technology on Surface Physics and Chemistry Laboratory, Jiangyou 621908, Sichuan, China}

\author{Xue-Bing Luo}
\affiliation{Science and Technology on Surface Physics and Chemistry Laboratory, Jiangyou 621908, Sichuan, China}

\author{Qiu-Yun Chen}
\affiliation{Science and Technology on Surface Physics and Chemistry Laboratory, Jiangyou 621908, Sichuan, China}

\author{Wei Feng}
\affiliation{Science and Technology on Surface Physics and Chemistry Laboratory, Jiangyou 621908, Sichuan, China}

\author{Qin Liu}
\affiliation{Science and Technology on Surface Physics and Chemistry Laboratory, Jiangyou 621908, Sichuan, China}

\author{Qun-Qing Hao}
\affiliation{Science and Technology on Surface Physics and Chemistry Laboratory, Jiangyou 621908, Sichuan, China}

\author{Yi Liu}
\affiliation{Science and Technology on Surface Physics and Chemistry Laboratory, Jiangyou 621908, Sichuan, China}

\author{Shi-Yong Tan}
\affiliation{Science and Technology on Surface Physics and Chemistry Laboratory, Jiangyou 621908, Sichuan, China}

\author{Xie-Gang Zhu}
\email{zhuxiegang@caep.cn}
\affiliation{Science and Technology on Surface Physics and Chemistry Laboratory, Jiangyou 621908, Sichuan, China}

\author{Hai-Feng Song}
\email{song\_haifeng@iapcm.ac.cn}
\affiliation{Laboratory of Computational Physics, Institute of Applied Physics and Computational Mathematics, Beijing 100088, China}
\affiliation{Software Center for High Performance Numerical Simulation,
China Academy of Engineering Physics,
Beijing 100088, China}

\author{Xin-Chun Lai}
\email{laixinchun@caep.cn}
\affiliation{Science and Technology on Surface Physics and Chemistry Laboratory, Jiangyou 621908, Sichuan, China}

\date{\today}

\begin{abstract}
The hybridization between the localized 4$f$ level ($f$) with conduction ($c$) states in $\gamma$-Ce upon cooling has been previously revealed in single crystalline thin films experimentally and theoretically, whereas its influence on the $\gamma\rightarrow\alpha$ phase transition was not explicitly verified, due to the fact that the phase transition happened in the bulk-layer, leaving the surface in the $\gamma$ phase. Here in our work, we circumvent this issue by investigating the effect of alloying addition of La on Ce, by means of crystal structure, electronic transport and ARPES measurements, together with a phenomenological periodic Anderson model and a modified Anderson impurity model. Our current researches indicate that the weakening of $f$-$c$ hybridization is the major factor in the suppression of $\gamma\rightarrow\alpha$ phase transition by La doping. The consistency of our results with the effects of other rare earth and actinide alloying additions on the $\gamma\rightarrow\alpha$ phase transition of Ce is also discussed. Our work demonstrates the importance of the interaction of $f$ and $c$ electrons in understanding the unconventional phase transition in Ce, which is intuitive for further researches on other rare earth and actinide metals and alloys with similar phase transition behaviors. 
\end{abstract}

\maketitle

\section{Introduction}

Among the $f$ electron systems, \emph{i.e.}, lanthanide and actinide metals, alloys and intermetallic compounds, cerium has its unique importance, in the sense that it is the first element possessing one $f$ electron and serves as the prototypical system in exploring the fascinating properties of $f$ electrons. Historically, cerium metal has been attracting the attentions of researchers, due to its complex phase transition under moderate pressure and temperature conditions, in which it is believed that its single $f$ electron has been playing important roles.  Face-centered cubic (FCC) $\gamma$-phase cerium will transform into the same FCC $\alpha$-phase, with a unconventional volume collapsed of up to 16.5\%, under a moderate pressure of $\sim$0.78 GPa at ambient temperature, or cooled down to low temperature ($\sim$141$\pm$10 K for bulk materials\cite{Koskenmaki:1978aa}, or down to around 50 K for single crystalline thin films\cite{Zhu:2020aa}) at ambient pressure.  As for the explanation of the mechanism of this intriguing phase transition, among the various theoretical models, $i.e.$, the promotional model\cite{Coqblin:1968aa,Ramirez:1971aa,Hirst:1974aa}, the Mott transition model\cite{Johansson:1974ab} and the Kondo volume collapse (KVC) model\cite{Allen:1982aa,Allen:1992aa},  the last one seems to be more adequate.  In the KVC model, it is believed that the increase of hybridization between $f$ electrons and conduction electrons drives the $\gamma\rightarrow\alpha$ phase transition in Ce metal,  and strong hybridization between $f$-$c$ has been revealed in the band structures of single crystalline Ce thin films characterized by high resolution angle resolved photoemission spectroscopy (ARPES) experiments\cite{Chen:2018aa,Zhu:2020aa,Wu:2021aa}.  However, direct experimental observation of the band dispersions of $\alpha$-Ce has not been achieved, due to the incompleteness of the $\gamma\rightarrow\alpha$ phase transition and persistent existence of $\gamma$ phase on the surface of Ce thin films, which makes the identification of $f$-$c$ hybridization evolution and the role it played across the $\gamma\rightarrow\alpha$ phase transition ambiguous. Therefore, approaches that could manipulate the $f$-$c$ hybridization and investigate its effects on the $\gamma\rightarrow\alpha$ phase transition are demanded.

For the purpose of determining the relationship between phase (atomic arrangement, basically) and electronic properties, pressure is one important variable\cite{Drickamer:1963aa}. It is plausible that only moderate hydrostatic pressure is needed to manipulate the $\gamma\rightleftarrows\alpha$ phase transitions in Ce, however, photoemission is technically non-applicable for hydrostatic pressurized solids. Another commonly adopted variable is alloying doping, which has been usually termed as chemical pressure. The effects on the $\gamma\rightleftarrows\alpha$ phase transition of Ce by doping with some rare earth and actinide elements have been studied by dilatometric and X-ray techniques\cite{Gschneidner:1962ab,Gschneidner:1962ac}. However, studies on the evolution of the electronic structure are still lacking. Among the various alloying dopants, La is one of a kind, with the reasons as follows: (1) with non-$f$ electron occupancy,  La acts as nonmagnetic ion, which could effectively tune the concentration of the localized $f$ electron momentum of Ce; (2) Ce and La alloyed in arbitrary chemical ratio could maintain the FCC structure in thin films; 
(3) La doping expands the FCC lattice as its concentration increases, from 5.16 \AA\ for Ce\cite{Alexander:2012aa} to 5.31 \AA\ for La\cite{Syassen:1975tw},  resulting in an effective negative chemical pressure. Here in our work, we have successfully grown Ce$_{1-x}$La$_x$ thin films by MBE, with La concentration $x$ being tuned continuously from zero to 100\%. Lattice structure characterization reveals that the thin films were in the FCC structure and the effective negative pressure was realized by La doping. Suppression of $\gamma\rightarrow\alpha$ phase transition by La doping was directly confirmed by electronic transport measurements. High resolution band structures of Ce$_{1-x}$La$_x$ thin films were obtained by ARPES, confirming that the electronic structure of 4$f$ electron in Ce could be effectively modified by La doping.  The evolution of $f$-$c$ hybridization by La doping was investigated by both a phenomenological periodic Anderson model (PAM) and a modified Anderson impurity model (the so-called GS model).  We show that the change in $f$-$c$ hybridization correlates with the suppression of $\gamma$-$\alpha$ phase transition in Ce, which is consistent with the scenario of KVC model.  

\section{Experimental methods}

Ce$_{1-x}$La$_x$ thin films were grown by molecular beam epitaxy (MBE). High purity Ce and La metals (99.9\%) were used and thoroughly degassed at 1600 $^\circ$C and 1500 $^\circ$C separately before thermal evaporation. La concentration $x$ was controlled by finely tuning the flux ratio of Ce and La during the thermal evaporation. Quartz crystal micro-balance was used to calibrate the flux rate of Ce and La down to an accuracy within 0.002 \AA/s. Single crystalline Ce$_{1-x}$La$_x$ thin films were deposited on Graphene/6H-SiC(0001) substrate for \emph{in-situ} band structure characterization,  while polycrystalline thin films were grown on Al$_2$O$_3$(11$\bar{2}$0) substrate for \emph{ex-situ} XRD and electronic transport measurements. For the photoemission spectroscopy (PES) experiment, the photoelectrons were excited by the He I$\alpha$ (21.2 eV) resonance line of a commercial Helium gas discharge lamp. The light was guided to the analysis chamber by a quartz capillary. In virtue of the efficient three-stage differential pumping system, the pressure in the analysis chamber was better than 1$\times$10$^{-10}$ mbar during our experiments, satisfying the harsh criteria for probing the physical properties of chemically high reactive elements. A VG Scienta R4000 energy analyzer was used to collect the photoelectrons. The total energy and momentum resolutions were better than 10 meV and 0.006 \AA$^{-1}$, respectively. The transport measurements were performed with the standard four-probe technique using a physical property measurement system (PPMS-9). Pure Indium (5N) metal was used to make the electrode contacts. The resistivity versus temperature curves were recorded during a cooling and warming cycle,  respectively.  It should be noted that different samples with different La concentrations were used in the above experiments to cover a wide range of La concentration.  Theoretical methods will be discussed in more detail in the following main context. 

\section{Results}

\subsection{Lattice structure and transport properties}
\begin{figure*}
\centering
\includegraphics[width=1.0\textwidth]{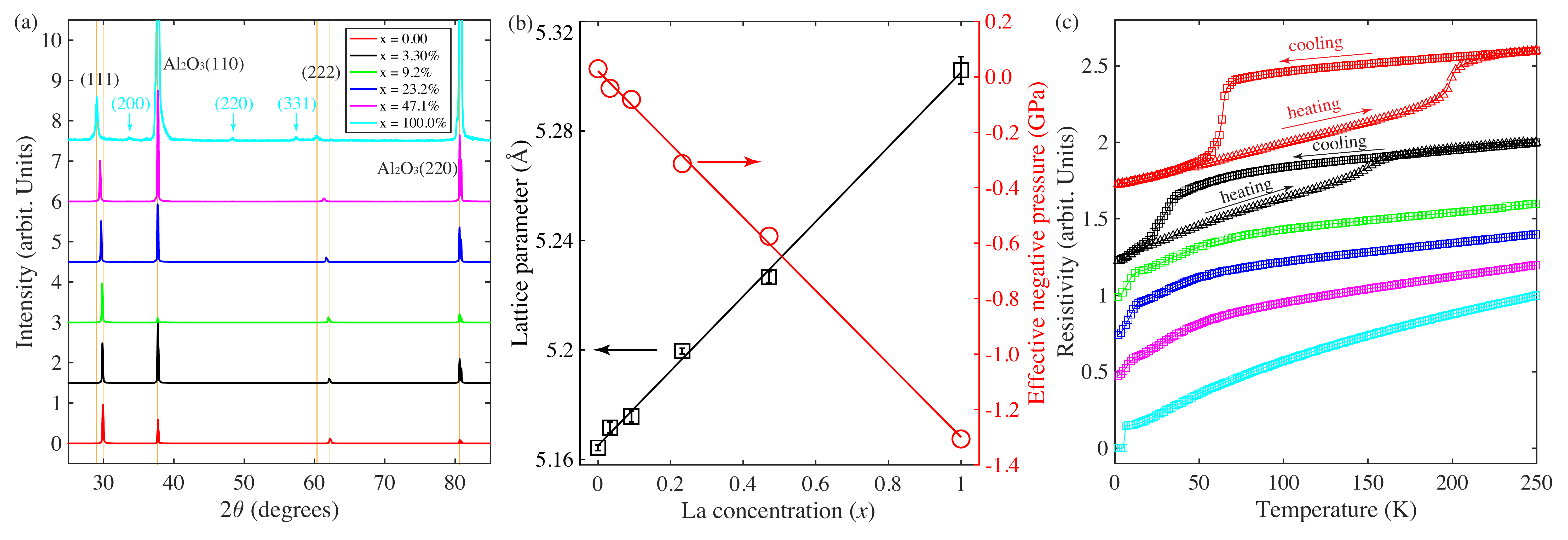}
\caption{(Color online) (a) Renormalized XRD curves of polycrystalline Ce$_{1-x}$La$_x$ thin films; (b) the evolution of lattice parameters in the FCC structure and the effective negative pressures by La doping; (c) normalized resistivity curves. }
\label{fig.XRDandRT}
\end{figure*}

Polycrystalline Ce$_{1-x}$La$_x$ samples with thickness of $\sim$ 1 $\mu m$ were used for transport measurement. XRD analysis was conducted just after the samples were taken out of the MBE chamber, as demonstrated in Fig. \ref{fig.XRDandRT}(a). The values of 2$\theta$ have been calibrated by the (110) and (220) peaks of the Al$_2$O$_3$(11$\bar{2}$0) substrate, and the intensities have been renormalized by the intensity of pure Ce (111) peak for a better view. According to the standard PDF cards, we know the Ce film (red curve in \ref{fig.XRDandRT}(a)) correspond to $\gamma$-Ce of FCC crystal structure.  The strong intensity of the (111) peaks for all the samples except pure La film indicates that the samples were highly oriented in the [111] crystallographic direction, although the \emph{in-situ} reflected high-energy electron diffraction (RHEED) showed all the films grown on Al$_2$O$_3$(11$\bar{2}$0) substrates were polycrystalline, which means that the films were mainly consisted with crystallites oriented along [111] out of plane and randomly \emph{in-plane}. The case for pure La film was quite different: the (111) was much weaker relative to that of the substrate, and the signal to noise ratio was much lower than that of Ce film, indicating that the crystallites in La film were totally randomly oriented.  The evolution of the lattice parameters could be seen from the continuous shift of the (111) and (222) peaks between the vertical lines in Fig. \ref{fig.XRDandRT}(a),  and in more detail in Fig. \ref{fig.XRDandRT}(b), demonstrating the linear expansion of the lattices by La doping. It should be stressed that the lattice parameters for our pure Ce and La thin films, \emph{i.e.}, 5.166$\pm$0.002 \AA\ and 5.307$\pm$0.005 \AA, corresponded to their FCC structures, respectively\cite{Alexander:2012aa,Bigci:2010aa}. 

To investigate the negative pressure effect by La doping, we have linearly extrapolated the pressure to volume (P-V) relationship of pure Ce to negative pressure\cite{Decremps:2011aa}, where the P-V relationship was almost linear for pressures lower than $\sim$0.70 GPa.  Based on our experimental lattice parameters, the effective negative pressures for various La concentrations were calculated and shown in Fig. \ref{fig.XRDandRT}(b).

The resistivity of the above Ce$_{1-x}$La$_x$ samples were renormalized by their values at 250 K and shown in Fig. \ref{fig.XRDandRT}(c). Hysteresis loop was observed in pure Ce and Ce$_{0.967}$La$_{0.033}$ thin films, which was the characteristic feature of the $\gamma\rightleftarrows\alpha$ phase transition. The area enclosed by the hysteresis loop shrank a lot by 3.30\% La doping, and diminished for La concentration at 9.20\% and above, which means that the $\gamma$-$\alpha$ phase transition was suppressed by a few percentage of La doping, \emph{i.e.}, less than 10\% of La concentration.  It should be noted that the lattice expansion for $x$ less than 10\% is smaller than $\sim$0.2\%, which is almost negligible. Thus, the mechanism of the suppression of the phase transition lies somewhere else, where the electronic properties of the $f$ electrons and their interaction with conduction electrons might be playing important roles.

\subsection{Photoemission Studies}
\begin{figure*}
\centering
\includegraphics[width=1.0\textwidth]{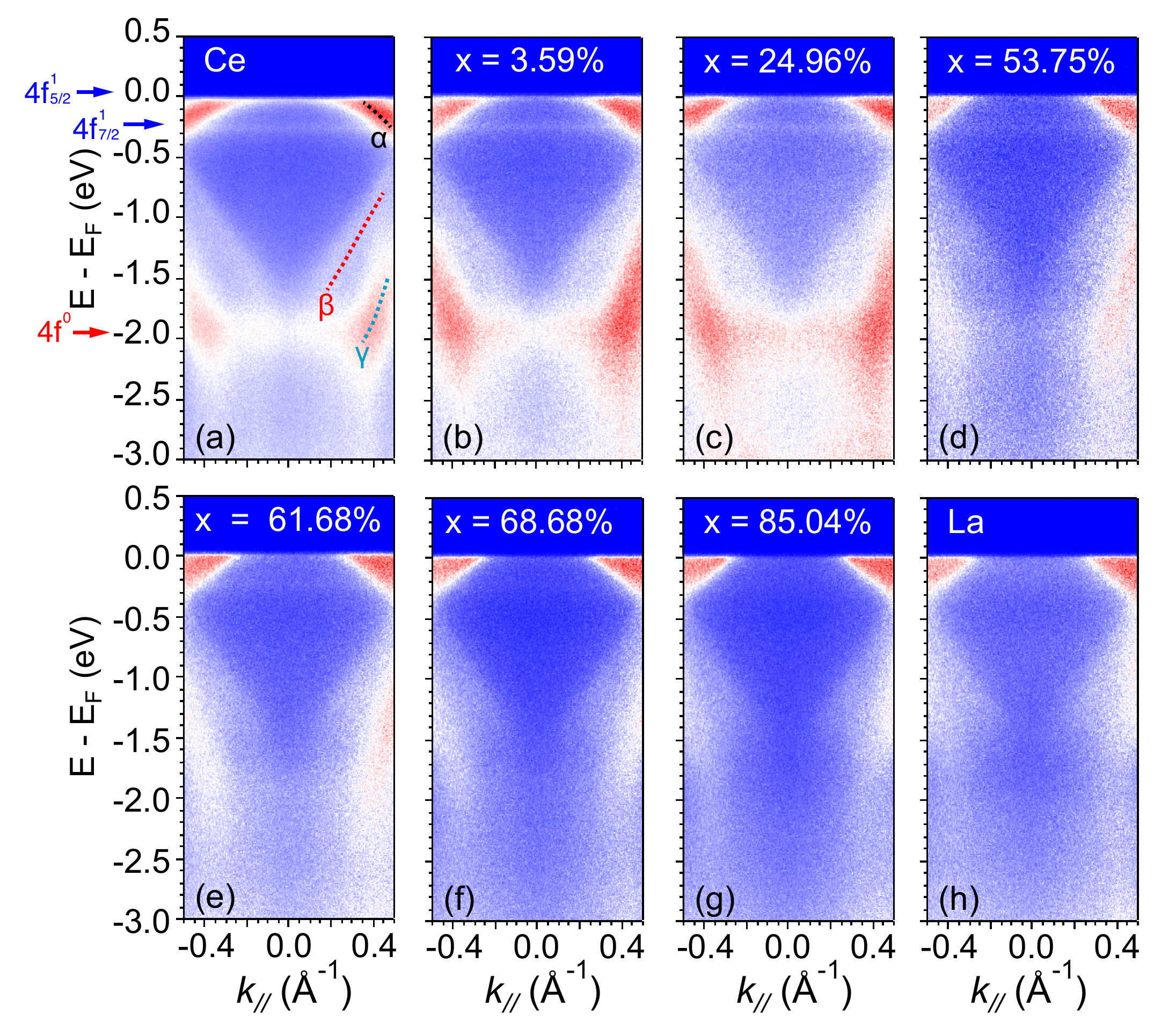}
\caption{(Color online) (a) - (h) ARPES spectra within large energy scale (-3.0 $\sim$ 0.5 eV) for Ce$_{1-x}$La$_x$ alloy thin films along $\bar{\rm{K}}$-$\bar{\Gamma}$-$\bar{\rm{K}}$ in the surface Brillouin zone (SBZ) for various La concentration.}
\label{fig.kgk}
\end{figure*}

Detailed studies on the electronic structures of pure Ce thin films by ARPES could be found in Ref. \onlinecite{Chen:2018aa,Zhu:2020aa,Wu:2021aa}. The ARPES spectra for our \emph{as-grown} Ce$_{1-x}$La$_x$ thin films with various La concentrations are shown in Fig. \ref{fig.kgk}, corresponding to the spectra taken along $\bar{K}-\bar{\Gamma}-\bar{K}$ in the surface Brillouin zone (SBZ). Fig. \ref{fig.kgk}(a) shows the main features of the band structures of Ce thin film: three dispersive bands that are originated from non-$f$ electrons ($spd$), labeled as $\alpha$, $\beta$ and $\gamma$, respectively; three non-dispersive bands that locate near Fermi level ($E_F$), $\sim$ 250 meV and $\sim$ 2 eV below $E_F$, which were conventionally attributed to the $4f^1_{5/2}$, $4f^1_{7/2}$ and $4f^0$, respectively. The clear presence of the $4f^1$ levels near $E_F$ indicates the strong hybridization of conduction electrons with the $4f$-electron in Cerium, enhancing its spectral intensity near $E_F$. As for pure La thin film, the ARPES spectrum in Fig. \ref{fig.kgk}(h) shows almost identical dispersive bands as observed in Ce, with the non-dispersive $f$ bands missing, which is a consequence of the absence of $f$-electron in La. With increasing La concentration in Ce$_{1-x}$La$_x$, the $4f$ bands evolved continuously and the conduction bands remained similar, as shown from Fig. \ref{fig.kgk}(b) to \ref{fig.kgk}(g).

\begin{figure*}
\centering
\includegraphics[width=1.0\textwidth]{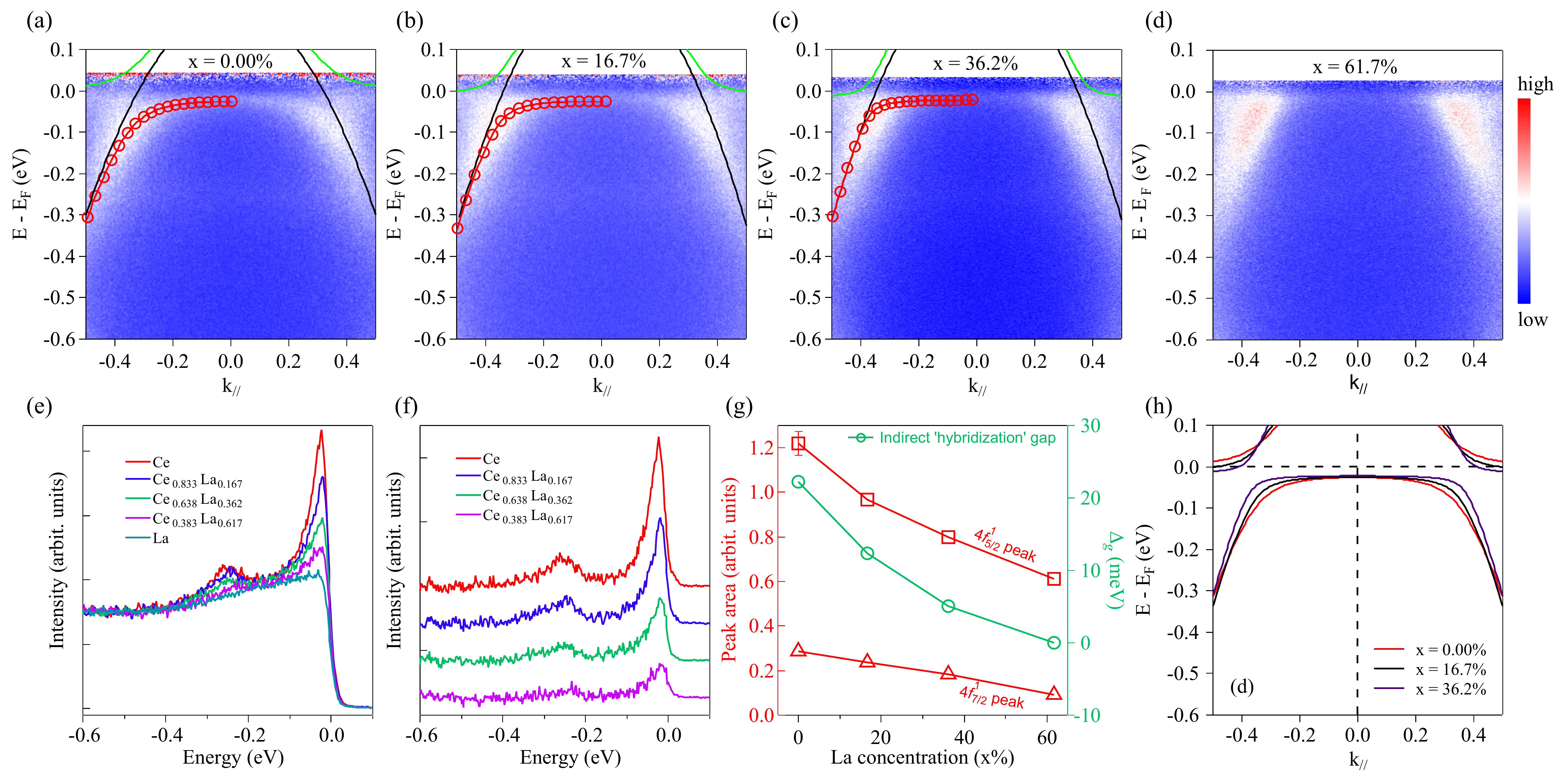}
\caption{(Color online) (a)-(d) Detailed band dispersions of Ce, Ce$_{0.833}$La$_{0.167}$, Ce$_{0.638}$La$_{0.362}$ and Ce$_{0.383}$La$_{0.617}$ thin films along $\bar{\rm{K}}$-$\bar{\Gamma}$-$\bar{\rm{K}}$, in which the $f$-$c$ hybridization bands are fitted by PAM model. The red dotted lines and green lines are the fitted results, while the black lines are the conduction band dispersions without $f$-$c$ hybridization. All the PAM fitting results are also summarized in (h). (e) Normalized energy distribution curves (EDCs) within $\pm$0.04 \AA\ around the $\bar{\Gamma}$ point of the thin films in (a) - (d), together with that of pure La thin film (ARPES spectra not shown here). (f) Normalized EDCs of (a) - (d) in (e) subtracted that of pure La thin films. The spectra are vertically shifted for a better view. (g) Peak areas of $4f^1_{5/2}$ (open squares) and $4f^1_{7/2}$ (open triangles) fitted from (f), together with the fitted indirect `hybridization' gaps $\Delta_g$ (open circles).}
\label{fig.edc}
\end{figure*}

To investigate the $k$-dependent $f$-$c$ hybridization, we introduce the phenomenological periodic Anderson model (PAM), which gives the band dispersion describing the hybridization as \cite{Chen:2017aa,Zhu:2020aa},
\begin{equation}
E_k^{\pm}=\frac{\epsilon_0+\epsilon(k)\pm\sqrt{(\epsilon_0-\epsilon(k))^2+4|V_k|^2}}{2}.
\label{fit.pam}
\end{equation}
Here $\epsilon_0$ is $4f$ ground state energy, $\epsilon(k)$ is the valence band dispersion at high temperature, and $V_k$ is the renormalized hybridization strength. When the La concentration $x$ is below 61.7\%, all band structures around the Fermi level of the Ce-La alloys can be fitted with the PAM, as shown in Fig. \ref{fig.edc}(a-c). Without regard to the $f$-$c$ hybridization, only conduction band crosses the Fermi level. When the hybridization strengthens, the conduction band bends to another directions with much smaller Fermi velocities and this is how the heavy electrons emerge. However, we find the $f$-$c$ hybridization gap of the Ce-La alloys decreases as the La concentration increases as shown in Fig. \ref{fig.edc}(h) and the green marker line in Fig. \ref{fig.edc}(g), indicating the decreased hybridization strength. Furthermore, when the La concentration is 61.7\% or more, as shown in Fig. \ref{fig.edc}(d), the valence band structure cannot be fitted with the PAM, namely the hybridization effect disappears, mainly resulted from the faint $f$ bands. That is to say, the hybridization strength of the Ce-La alloys monotonously decreases as the La concentration increases until its disappearance. 

Additionally, we also analyzed the evolution of the $f$ band intensities by La doping, which was done by investigating the energy distribution curves (EDCs) of the spectra. 
The EDC peak intensities of both $4f^1_{5/2}$ and $4f^1_{7/2}$ bands gradually decrease as the La concentration increases as shown in Fig. 3(e) and (f). This phenomenon is quite similar to the $f$ band evolutions of Ce film as a function of temperature and mainly results from the localization of $f$ electrons with increasing temperature. We can also find that the $4f^1_{5/2}$ and $4f^1_{7/2}$ peak areas of the EDCs decrease with increasing La concentration as shown in Fig. 3(g). When the $f$ electron concentration is extremely low (see EDC of Ce$_{0.383}$La$_{0.617}$ in Fig. \ref{fig.edc}(e) and (f)), there still have some $4f^1_{5/2}$ intensities around the Fermi level. However, the $f$-$c$ hybridization does not happen and the band dispersion cannot be fitted with the PAM as shown in Fig. \ref{fig.edc}(d). With the increasing $f$ electron concentration, the intensity of $f$ band strengthens and the $f$-$c$ hybridization also emerges as shown in Fig. \ref{fig.edc}(a)-(c). 


\subsection{Theoretical consideration}
From the above photoemission results, we could see the persistent existence of $f$-$c$ hybridization in Ce$_{1-x}$La$_x$ at low temperature for La concentrations over quite a large range. And the hybridization effect could be well demonstrated by the phenomenological periodic Anderson model, even La doping could have driven the 4$f$ electrons from the Kondo lattice to the single impurity configuration. Therefore, it is intriguing and informative to investigate the electronic behaviors of Ce$_{1-x}$La$_x$ alloys in the impurity model. In the work of Gunnarsson and Sch{\"o}nhammer, the electron spectroscopies of Ce related compounds, including the x-ray photoemission, 3$d$$\rightarrow$4$f$ x-ray absorption, the valence photoemission, and the bremsstrahlung isochromat spectra, have been calculated in a modified Anderson impurity model\cite{Gunnarsson:1983ab}. This so-called GS model has been well verified in its applications in a series of Ce compounds\cite{Gunnarsson:1983aa,Patthey:1990aa}. Here in our work, we have applied the GS model to simulate the photoemission spectra of our samples. To simulate the three peaks structure (4$f^0$, 4$f^1_{7/2}$ and 4$f^1_{5/2}$) of the photoemission spectra of Ce related materials, the spin-orbit splitting of the $f$ level was include in the GS model, and the resultant valence photoemission spectra could be described analytically in the first-order approximation as follows,

\begin{widetext}
\begin{equation}
\rho_v(\epsilon)=A^2\int d\epsilon^\prime \vert V(\epsilon^\prime) \vert^2 \left[ \frac{N_{f1}}{(\Delta E-\epsilon_f+\epsilon^\prime)^2}+\frac{N_{f2}}{(\Delta E-\epsilon_f-\Delta\epsilon_f+\epsilon^\prime)^2} \right]\widetilde{g}(z-\Delta E+\epsilon_f-\epsilon^\prime)
\label{gs.pes}
\end{equation}
\end{widetext}

where $V(\epsilon)$ describes the hopping between the $f$ level and the conduction states, and
is related to the coupling strength $\Delta$ by $\Delta=\pi \text{max}_\epsilon[|V(\epsilon)|^2]$. $N_{f1}$ and $N_{f2}$ are the degeneracy of the $j=\frac{7}{2}$ and $j=\frac{5}{2}$ levels, \emph{i.e.}, 8 and 6 respectively. $\Delta E$ is the lowering of the energy when the $f$ electron impurity is introduced, and could be calculated numerally. $\epsilon_f$ and $\Delta\epsilon_f$ are the position and the spin-orbit splitting of the $f$ level, respectively. The definition of $\widetilde{g}$ could be found in the appendix of the literature\cite{Gunnarsson:1983ab}.  To numerically calculate $\rho_v(\epsilon)$, an infinitesimal $0^+$ was introduced ($z=\epsilon-i0^+$) in the integration procedure, and the imaginary part of the integration corresponds to the valence photoemission spectra. 

As for $|V(\epsilon)|^2$,  instead of using the over-simplified semi-elliptical form\cite{Gunnarsson:1983ab}, we have adopted a double-peaked Lorentzian line-shape to account for the coupling between the $f$ level and different conduction bands. In our simulation, the background includes two terms: the first is a quadratic part, which corresponds to the systematic noise and second order photoelectrons; the second is the contribution from conduction bands, and is related to the linear combination of the Lorentzian peaks in $|V(\epsilon)|^2$.  During the fitting procedure, the spin-orbit splitting $\Delta\epsilon_f$ was fixed at 280 meV\cite{Buchanan:1966aa,Patthey:1985aa}. The starting values of $\Delta$ and $\epsilon_F$ were 0.125 eV and -1.2 eV, respectively. A Fermi-Dirac distribution (FDD) of 80 K was also introduced, which corresponds to the experimental temperature. All the fitting parameters were constrained in reasonable ranges in order to guarantee that the results have physical meaning. 

\begin{figure*}
\centering
\includegraphics[width=1.0\textwidth]{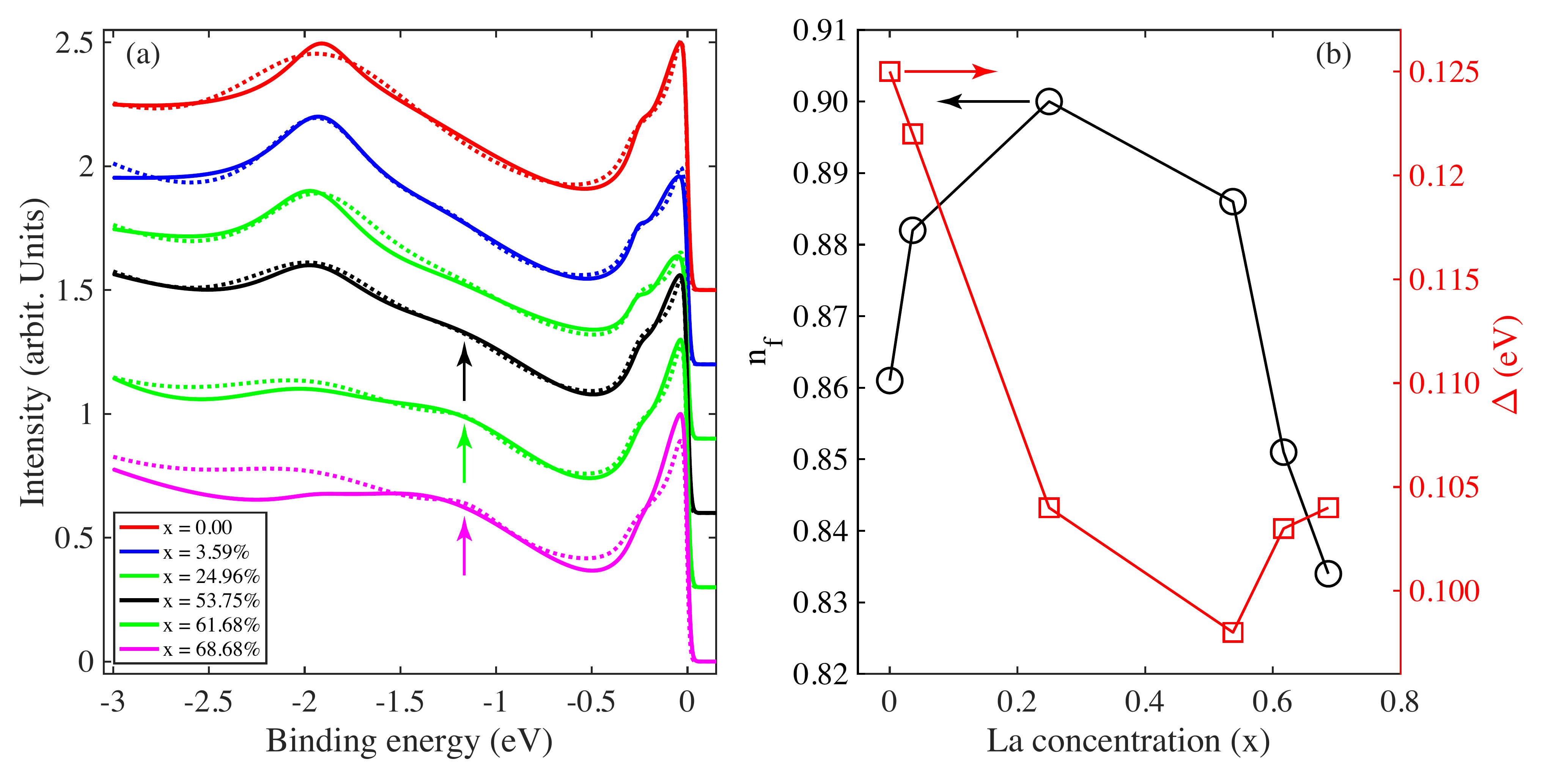}
\caption{(Color online) (a) Experimental (solid lines) and fitted (dotted lines) valence photoemission spectra of Ce$_{1-x}$La$_x$ thin films; (b) Calculated $f$ electron occupancy $n_f$ and fitted coupling strength $\Delta$ for various La concentration.}
\label{fig.fit}
\end{figure*}

The experimental valence spectra were extracted from Fig. \ref{fig.kgk} by integrating the EDCs within $\pm$0.50 \AA$^{-1}$ around the $\bar{\Gamma}$ point, and were further smoothed by the multicomponent function as mentioned in Ref. \onlinecite{Zhu:2020aa}. The resultant smoothed spectra were demonstrated as solid lines in Fig. \ref{fig.fit}(a). The dotted lines in Fig. \ref{fig.fit} were the fitted valence spectra by the GS model.  The overall fittings are quite satisfactory for La concentration below 53.75\%,  where the location, relative intensities and shape of the 4$f^0$ ionization peak,  the Kondo peak near $E_F$ and its 4$f^1_{7/2}$ spin-orbit splitting duplicate were well reproduced. As La concentration increases further,  the spectral intensity of the normal conduction band feature (indicated by the vertical arrows in Fig.  \ref{fig.fit}(a) gradually prevails that of the 4$f^0$ peak, and the fitting results deviate from the experimental spectral below -1.5 eV.  The calculated $f$ electron occupancy for pure Ce is approximate 0.86, and increases to $\sim$0.90 for $x$ = 24.06\%, as shown in Fig. \ref{fig.fit}(b). The increase of $n_f$ indicates that the $f$ electrons in Ce$_{1-x}$La$_x$ become more localized, or in other word, the coupling strength of the $f$ level with the conduction states weakens, which is consistent with the previous discussed PAM fitting results of the $f$-$c$ hybridization. However, the relationship between $n_f$ and $x$ changes as $x$ goes beyond 24.96\%, \emph{i. e.}, $n_f$ decreases monotonically as $x$ increases, as shown in Fig. \ref{fig.fit}(b). This could be attributed to the fact that the valence spectra demonstrate more and more La features as $x$ increases, and Ce atoms act as dopant instead of host, in which cases the GS model seems to be not adequate.  Similar behaviors were also observed for the coupling strength $\Delta$, as shown by the red marker line in Fig. \ref{fig.fit}, despite that $\Delta$ decreased with La concentration up to 53.75\%.

As mentioned above, the central idea of the GS model originated from the impurity Anderson model, and the coupling of $f$-level with conduction states resulted in the 3-peak feature (4$f^0$, 4$f^1_{5/2}$ and 4$f^1_{7/2}$) in the valence spectra of many Ce related materials. For the validity of this model, the screening of the local momentum of $f$ electrons by the conduction electrons should be moderate, and lattice structure of Ce should be intact to ensure the formation of $f$ bands.  La doping, in on hand, would effectively increase the ratio of conduction electrons to $f$ electrons, \emph{i.e.}, strengthening the screening of $f$ electrons,  and on the other hand, would gradually destroy the original Ce lattice, both of which could drive the system from strong $f$-$c$ coupling to weak coupling, and eventually to no coupling.  Additionally, according to the previous XRD results, La doping gradually expands the FCC lattice, exerting effective negative pressure on the lattice, which would also weaken the hybridization between $f$ and conduction electrons, despite the fact that its effect is negligible for a few percentage of La doping. Therefore,  we could safely conclude that the suppression of $\gamma$-$\alpha$ phase transition is mainly due to the weakening of $f$-$c$ hybridization, and on the contrary, the strengthening of $f$-$c$ hybridization could drive the $\gamma\rightarrow\alpha$ phase transition upon the condition of cooling or positive pressure.


\newpage
\section{Discussions and Conclusions}
La doping in Ce has two major influences: diluting the $f$ electron concentration and expanding the lattice parameters. Our researches have demonstrated that the former would greatly weaken the $f$-$c$ hybridization effect, and consequently suppress the $\gamma$-$\alpha$ phase transition in Ce upon cooling, or in other words, stabilize the $\gamma$ phase.  This indicates that greater pressure should be expected for the $\gamma\rightarrow\alpha$ phase transition to take place for La doped Ce samples at room temperature, which was indeed the case from the experimental results on bulk samples\cite{Gschneidner:1962ac}.Additionally, many rare earth alloying additions on Ce, such as Sc, Pr, Dy and Lu, shrank the lattice by exerting positive chemical pressure and increased the phase transition pressure at room temperature\cite{Shunk:1969wm,Gschneidner:1962ab,Gschneidner:1962ac}. One special alloying addition is the actinide element plutonium, which also shrank the lattice from 0 up to $\sim$25\% concentration, whereas decreasing the transition pressure, \emph{i.e.}, Ce was prone to phase transition upon Pu doping. It is well known that Ce and Pu are counterparts to each other\cite{Moore:2009,Clark:2019aa}, in the sense that: on one hand, they both undergo huge volume collapse, 17\% for Ce and 25\% for Pu, across their $\gamma\rightarrow\alpha$ and $\delta\rightarrow\alpha$ phase transitions, respectively; and on the other hand, the 4$f$-electron of Ce and 5$f$-electron of Pu both have dual itinerant and localized behaviors, demonstrating strong many-body effects and strong interaction with their conduction electrons. We conjecture that Pu doping would strengthen the $f$-$c$ hybridization, whereas other alloying additions would not. Another interesting alloying addition is thorium, which also shrank the FCC lattice of Ce, however, the pressure needed for the $\gamma\rightarrow\alpha$ phase transition at room temperature only decreased slightly, with a value of $\sim$3.6\% at 20\% of Th doping, compared with the value of $\sim$71.5\% for 20\% of Pu doping\cite{Gschneidner:1962ab,Gschneidner:1962ac}.  What is more, the $\gamma\rightarrow\alpha$ phase transition persistently existed upon cooling for a large range of Th doping from 0 to 60\%. Despite the atomic ground state configuration 6$d^2$7$s^2$, Th metal in the FCC structure has a non-integer 5$f$-electron count due to the 5$f$-6$d$ overlapping\cite{Johansson:1995aa}, or in other word, $f$-$c$ hybridization. Alloying the two elements most probably did not change the overall valence states and $f$-electron occupancy too much, thus Th doping had much smaller influence on the phase transition of Ce, compared with the other alloying elements. 

To summarize, we have experimentally and theoretically investigated the effects of alloying addition on Ce by La. Lattice structure and electronic transport experiments have revealed linear expansion of the FCC lattice and suppression of the $\gamma\rightarrow\alpha$ phase transition. Detailed ARPES measurements and theoretically model calculations have demonstrated the weakening of $f$-$c$ hybridization and the important role it played in the suppression of the $\gamma\rightarrow\alpha$ structural phase transition.  Our results provoke further studies on the electronic origins of the phase transitions and phase stabilities of rare earth and actinide metals and alloys. 


\section{Acknowledgments}

The work was supported by the National Key Research and Development Program of China (2021YFA1601100, 2017YFA0303104), the SPC-Lab Research Fund (WDZC201901), the Science Challenge Project (TZ2016004 and TZ2018002), the National Science Foundation of China (U1630248, 11774320, 11904334), Special Funds of Institute of Materials (TP02201904) and the Development Funds (JZX7Y201901SY00900107). 

\bibliography{cela}


%

%

\end{document}